\documentstyle{article}

\catcode`\@=11
 
\@addtoreset{equation}{section}
\catcode`\@=12
\font\bbb=msbm10

\begin{document}

\title{Ablowitz-Ladik system with discrete potential. I. Extended resolvent}
\author{A.~K.~Pogrebkov\thanks{Steklov Mathematical Institute, Moscow, 
Russia; e-mail: pogreb@.mi.ras.ru}
and M.~C.~Prati\thanks{Scuola Normale Superiore di Pisa, INFN, Sezione di 
Pisa, ITALIA; e-mail: prati@sns.it}}
\maketitle

\begin{abstract}
Ablowitz-Ladik linear system with range of potential equal to \{$0,1$\} is
considered. The extended resolvent operator of this system is constructed
and the singularities of this operator are analyzed in detail.
\end{abstract}

\section{Introduction}

Our aim in this article is to study the spectral theory of the matrix 
operator $L(w)$, 
\begin{eqnarray}
L_{m,n}^{}(w) &=&\delta _{m,n-1}^{}-\left( 
\begin{array}{cc}
w & r_{n}^{} \\ 
s_{n}^{} & 1/w
\end{array}
\right) \delta _{m,n}^{},\qquad  \label{1} \\
m,n &\in &\hbox{\bbb Z},\qquad w\in \hbox{\bbb C},  \nonumber
\end{eqnarray}
every element of which is a $2\times 2$ matrix, $\delta _{m,n}^{}$ is the
Kronecker symbol and we omitted a $2\times 2$ unit matrix factor in the
term $\delta _{m,n-1}$. Our attention is concentrated to the case where
values of both potentials, $r_{n}$ and $s_{n}$, are equal to $0$ and $1$: 
\begin{equation}
r_{n},s_{n}\in \{0,1\},\qquad n\in \hbox{\bbb Z}.  \label{2}
\end{equation}
Moreover, we consider here the case of potentials with finite support, i.e.,
for every given potential there exist finite $k$ and $K$, $k\leq K$, 
$k,K\in\hbox{\bbb Z}$---lower and upper borders of the support---such that 
\begin{equation}
r_{n}^{}=s_{n}^{}=0,\qquad n\leq k-1,\qquad n\geq K+1.  \label{25}
\end{equation}

The corresponding linear problem, 
\begin{equation}
L(w)\Phi =0,  \label{3}
\end{equation}
is the Ablowitz--Ladik problem~[1,2] which is known to be a discretized
version of the Zakharov--Shabat linear problem. And like the latter the
Ablowitz--Ladik problem is associated to a variety of
differential--difference integrable equations, such as discrete mKdV
equation, difference KdV, Toda chain, etc., [3]. Problem~(\ref{3}) describes
also discrete systems with nonanalytic dispersion relations~[4].

The Ablowitz--Ladik problem is also known~[5,6] to be associated to
dif\-fe\-ren\-ce--difference nonlinear equations, that are related to some
class of cellular automata, i.e., dynamical systems in a discrete
space--time with values belonging to some finite field, say, 
$\hbox{\bbb F}_{2}$. Cellular automata attract great interest in the 
literature because of the wide range of their applications in different 
sciences, from physics to biology, from chemistry to social sciences. 
Detailed references for these applications can be found in~[7--11]. These 
automata are also subject to intensive mathematical study, see for 
example~[12--22]. It is just this kind of applications of problem~(\ref{3}) 
that motivated our specific choice of condition~(\ref{2}) on potential.

The problem of the investigation of~(\ref{3}) by means of the inverse 
scattering transform, as it was performed in [3], becomes obvious if we 
write down this equation explicitly: 
\begin{equation}
\Phi _{n+1}=\left( 
\begin{array}{cc}
w & r_{n} \\ 
s_{n} & 1/w
\end{array}
\right) \Phi _{n},\qquad n=0,\pm 1,\pm 2,\ldots.  \label{3'}
\end{equation}
In the standard approach to the study of the spectral problems, the main
objects of the theory---the Jost solutions---are determined by their
asymptotics at $n\rightarrow +\infty $ and $n\rightarrow -\infty $. A solution
given by its asymptotics at $n\rightarrow -\infty $ is swept from the left
by~(\ref{3'}). But in order to construct the Jost solution given by its
asymptotics at $n\rightarrow +\infty $, one has to invert the matrix in the
r.h.s.\ of~(\ref{3'}). The determinant of this matrix is equal to 
$1-r_{n}s_{n}$, so in the standard approach the condition $r_{n}s_{n}\neq 1$ 
must be fulfilled. In the case where the potential satisfies~(\ref{2}) this 
means that  for every $n$ either $r_{n}$ or $s_{n}$ must be equal to
zero~[6]. Such condition drastically restricts the class of potentials of
the type~(\ref{2}), so our aim in this and forthcoming publications is to
elaborate an extension of the inverse scattering transform method to the case
where both $r_{n}$ and $s_{n}$ can be equal to $1$. Let us also emphasize
that, imposing condition~(\ref{2}) on the potentials, we do not use here 
the condition $r_{n},s_{n}\in \hbox{\bbb F}_{2}$. As was speculated in~[23] 
the problem of the integrability of the cellular automata or, more 
precisely, the problem of existence of the Lax representations must be 
solved in terms of the exact equalities, and not in terms of equalities on 
some finite field.

The fact that some matrix(--matrix) operator $L$ is analogous to a
differential one is reflected in the property that matrix elements $L_{m,n}$ 
are different from zero only for uniformly bounded values of $|m-n|$. In the
case of~(\ref{1}) we have $1\geq m-n\geq 0$. Consequently, we can apply the 
resolvent approach~[24], [25] to investigation of the Ablowitz--Ladik 
problem. The preliminary results of our investigation were published 
in~[26]. 

The resolvent approach is based on the following extension of the operator 
$L(w)$: 
\begin{equation}
L_{m,n}^{}(w,h)=h_{}^{n-m}L_{m,n}^{}(w),  \label{4}
\end{equation}
where $h$ is a real non-negative parameter. In particular for the operator 
(\ref{1}) we have 
\begin{equation}
L_{m,n}^{}(w,h)=h\delta _{m,n-1}^{}-u_{n}^{}(w)\delta _{m,n}^{},  \label{5}
\end{equation}
where we introduced 
\begin{equation}
u_{n}^{}(w)=\left( 
\begin{array}{rr}
w & r_{n}^{} \\ 
s_{n}^{} & 1/w
\end{array}
\right)\equiv w^\sigma_{}+ \left( 
\begin{array}{rr}
0 & r_{n}^{} \\ 
s_{n}^{} & 0
\end{array}
\right) ,\qquad n\in \hbox{\bbb Z},  \label{6}
\end{equation}
and $\sigma$ is the Pauli matrix $\sigma_3$, 
\[
\sigma=\sigma_3^{}= \left( 
\begin{array}{rr}
1 & 0 \\ 
0 & -1
\end{array}
\right). 
\]

If we have some infinite matrix--matrix operator $A_{m,n}(h)$ depending on
a parameter $h$ we can associate to it the Laurent series
\begin{equation}
A(\zeta,\zeta',h)=\sum_{m,n=-\infty }^{+\infty }\!\zeta_{}^{-m} 
{\zeta'}^{n}A_{m,n}^{}(h),\qquad \zeta,\zeta'\in 
\hbox{\bbb C},\qquad |\zeta|=|\zeta'|=1.  \label{7}
\end{equation}
In what follows we consider matrices $A_{m,n}(h)$ such that the 
series~(\ref{7}) are convergent in the sense of Schwartz distributions in 
$\zeta$, $\zeta'$ ($|\zeta|=|\zeta'|=1$) and $h$ ($h\geq 0$). The elements 
$A_{m,n}(h)$ are reconstructed by means of the formula
\begin{equation}
A_{m,n}(h)=\oint\limits_{|\zeta|=1}\frac{d\zeta\, \zeta_{}^{m-1}}{2\pi i}
\oint\limits_{|\zeta'|=1}\frac{d\zeta'\,\zeta_{}^{\prime -n-1}}
{2\pi i}A(\zeta,\zeta',h).  \label{8}
\end{equation}

In order to explain the meaning of the extension~(\ref{4}) let us introduce 
the function
(distribution) 
\begin{equation}
A(\zeta,z)=A(\zeta e_{}^{i\arg z},e_{}^{i\arg z},|z|)  \label{9}
\end{equation}
where $\zeta,z\in \hbox{\bbb C}$, $|\zeta|=1$; by~(\ref{4}) and~(\ref{7}) 
\begin{equation}
A(\zeta,z)=\sum_{m,n}\!z_{}^{n-m}\zeta_{}^{-m}A_{m,n}^{}.  \label{10}
\end{equation}
Then the above mentioned similarity of matrix and differential operators
means that $A(\zeta,z)$ depends on $z$ and $z^{-1}$ polynomially. Let us
mention that if we have two objects of this kind, $A$ and $B$, their product
(composition) is defined as follows:
\begin{eqnarray}
&&(AB)_{m,n}^{}(h) =\sum_{l=-\infty }^{+\infty }A_{m,l}^{}(h)B_{l,n}^{}(h),
\label{a11} \\
&&(AB)(\zeta,\zeta'',h) =\oint\limits_{|\zeta'|=1} 
\frac{d\zeta'\,\zeta_{}^{\prime -n-1}}{2\pi i}A(\zeta,\zeta',h)
B(\zeta',\zeta'',h),
\label{12} \\
&&(AB)(\zeta,z) =\oint\limits_{\,|\zeta'|=1}\!\frac{d\zeta'}
{2\pi i\zeta'}A(\zeta\overline{\zeta'},z\zeta')B(\zeta',z),  \label{13}
\end{eqnarray}
where the left hand sides of these equations are related through 
of~(\ref{7})--(\ref{10}). The main object of our investigation is inverse 
$M(w)$ of the operator $L(w)$ extended by~(\ref{4}), 
\begin{equation}
L(w)M(w)=I,\qquad M(w)L(w)=I.  \label{14}
\end{equation}
In matrix notations the first equality thanks to~(\ref{5}) has the form
\begin{equation}
hM_{m+1,n}^{}(w,h)=\delta _{m,n}^{} +u_{m}^{}(w)M_{m,n}^{}(w,h).  \label{15}
\end{equation}

In order to define this inversion in a unique way we introduce

{\bf Definition 1.} A solution $M(w)$ of~(\ref{14}) is called extended
resolvent of the operator $L(w)$ if $M(w,\zeta,\zeta',h)$ is a Schwartz
distribution with respect to $\zeta$ and $\zeta'$ and a sectionally 
continuous function of $h$, $h\geq 0$.

Let us first consider the case of zero potential, i.e., 
$r_{n}\equiv s_{n}\equiv 0$. Then the resolvent which we denote by 
$M_{0}(w)$ obeys the following equation
\begin{equation}
hM_{0,m+1,n}^{}(w,h)=\delta _{m,n}^{}+w_{}^{\sigma }M_{0,m,n}^{}(w,h).
\label{16}
\end{equation}
It is convenient to rewrite this equation using representation~(\ref{10}): 
\begin{equation}
\left( \zeta z-w^{\sigma }\right) M_{0}^{}(w,\zeta,z)=\delta _{c}(\zeta-1),
\label{17}
\end{equation}
where we introduced the $\delta $-function on $|\zeta|=1$, 
\begin{equation}
\delta _{c}^{}(\zeta-1)=\sum_{n=-\infty }^{\infty }\zeta_{}^{n},
\label{18}
\end{equation}
so that 
\begin{equation}
\oint\limits_{|\zeta|=1}\frac{d\zeta}{2\pi i}\delta _{c}^{}(\zeta
-1)f(\zeta)=f(1)  \label{19}
\end{equation}
for an arbitrary test function $f(\zeta)$ on the contour. Then 
\begin{equation}
M_{0}(w,\zeta;z)=(\zeta z-w^{\sigma })_{}^{-1}\delta _{c}(\zeta-1),
\label{20}
\end{equation}
so that by~(\ref{9}) 
\begin{equation}
M_{0}(w,\zeta,\zeta',h)=(\zeta h-w^{\sigma })_{}^{-1}\delta
_{c}(\zeta/\zeta'-1),  \label{201}
\end{equation}
or by~(\ref{8}) 
\begin{eqnarray}
&&M_{0,m,n}(w,h) =h_{}^{n-m}w_{}^{\sigma(m-n-1)}\Bigl\{\theta(h-|w^{\sigma}|)
\theta (m\geq n+1)-  \nonumber \\
&&\qquad -\theta (|w^{\sigma }|-h)\theta (n\geq m)\Bigr\},  \label{21}
\end{eqnarray}
where we introduced the matrices 
\begin{eqnarray}
\theta (h-|w^{\sigma }|) &=&\left( 
\begin{array}{rr}
\theta (h-|w|) & 0 \\ 
0 & \theta (h-1/|w|)
\end{array}
\right) ,  \nonumber \\
&&  \nonumber \\
\theta (|w^{\sigma }|-h) &=&\left( 
\begin{array}{rr}
\theta (|w|-h) & 0 \\ 
0 & \theta (1/|w|-h)
\end{array}
\right) .  \label{a22}
\end{eqnarray}
Here we have to make some comments. First, by~(\ref{4}), all expressions 
$h^{m-n}L_{m,n}(w,h)$ are independent on $h$ and equal to $L_{m,n}(w)$, 
see~(\ref{1}). On the contrary, $h^{m-n}M_{0,m,n}(w,h)$ essentially depends 
on $h$ and it is just this dependence that guaranties that 
$M_{0}(w,\zeta,\zeta',h)$ exists as a distribution in $\zeta$, $\zeta'$. 
Second, any solution of the homogeneous equation $L_{0}(w)M_{0}(w)=0$ is 
proportional to $\delta (h-|w^{\sigma }|)$, where the matrix 
$\delta$-function is defined in analogy with~(\ref{a22}). Thus we see that
the condition set on $M_{0}(w,\zeta,\zeta',h)$ in Definition~1 to be
a sectionally continuous function of $h$ enables us to define the resolvent 
$M_{0}(w)$ uniquely. In what follows we consider the case of a nontrivial 
potential satisfying conditions~(\ref{2}) and~(\ref{25}).

\section{Extended resolvent of the regularized operator}

The specific problem connected with equation~(\ref{15}) is, as was mentioned 
above in the discussion of Eq.~(\ref{3}), that if $r_{n}=s_{n}=1$ the matrix 
$u_{n}(w)$ is not invertible. Thus, first of all we have to introduce some
regularization of $u_{n}(w)$, say, 
\begin{eqnarray}
&&u_{n}^{}(w)\rightarrow u_{n}^{}(w,\lambda ) =\left( 
\begin{array}{rr}
(\lambda r_{n}^{}s_{n}^{}+1)w & r_{n}^{} \\ 
s_{n}^{} & 1/w
\end{array}
\right) =  \nonumber \\
&&\qquad =\lambda (1-\det u_{n}^{}(w)){\frac{1+\sigma }{2}}+u_{n}^{}(w).  
\label{23}
\end{eqnarray}
This substitution regularizes singular only $u_{n}$ (i.e., such that 
$\det u_{n}=0 $), leaving all other $u_{n}$ untouched. Indeed, by~(\ref{2}) 
$\det u_{n}$ equals either $0$ ot $1$. Then 
\begin{equation}
\det u_{n}^{}(w,\lambda )=\left\{ 
\begin{array}{cc}
1, & \det u_{n}^{}(w)=1, \\ 
\lambda , & \det u_{n}^{}(w)=0.
\end{array}
\right.  \label{24}
\end{equation}

Thus we start with the regularized operator 
\[
L_{m,n}(w,\lambda,h)=h\delta_{m,n-1}-u_m(w,\lambda)\delta_{m,n} 
\]
(cf.~(\ref{54})), i.e., by~(\ref{23}) 
\begin{equation}
L(w,\lambda)=L(w)-\lambda D,  \label{240}
\end{equation}
where we introduced the diagonal operator 
\begin{equation}
D_{m,n}^{}=\frac{1+\sigma }{2}(1-\det u_{n}^{})\delta _{m,n}^{}.  \label{244}
\end{equation}
Correspondingly, we denote the extended resolvent of the regularized
operator as $M(w,\lambda )$. It obeys equations (cf.~(\ref{14})) 
\begin{equation}
L(w,\lambda )M(w,\lambda )=I,\qquad M(w,\lambda )L(w,\lambda )=I,
\label{241}
\end{equation}
that by means of~(\ref{240}) can be written in the form 
\begin{eqnarray}
&&[ L(w)-\lambda D] M(w,\lambda ) =I,  \label{242} \\
&&M(w,\lambda )[ L(w)-\lambda D] =I.  \label{243}
\end{eqnarray}
Properties of $M(w,\lambda )$ in the limit $\lambda\rightarrow 0$ are 
studied in the next section.

Let for simplicity write
\begin{equation}
\widetilde{M}_{m,n}^{}=h_{}^{m-n}M_{m,n}^{}(w,\lambda ,h),  \label{26}
\end{equation}
i.e. we omit for a while dependencies on $w$, $\lambda $, and $h$. Then 
Eq.~(\ref{242}) takes the form 
\begin{equation}
\widetilde{M}_{m+1,n}^{}=\delta _{m,n}^{}+u_{m}^{}\widetilde{M}_{m,n}^{},
\label{27}
\end{equation}
where dependence of $u_{m}$ on $w$ and $\lambda $ is also omitted. It is
easy to check that for any $m\geq m'$ we have from~(\ref{27}) 
\begin{equation}
\widetilde{M}_{m,n}^{}=\theta (m\geq n+1)\theta (n\geq m') 
\overleftarrow{\prod_{l=n+1}^{m-1}}u_{l}^{}+\left( \overleftarrow
{\prod_{l=m'}^{m-1}}u_{l}^{}\right) \widetilde{M}_{m',n}
^{},  \label{28}
\end{equation}
where we introduced the notation 
\begin{equation}
\theta (m\geq n)=\displaystyle\left\{ 
\begin{array}{cc}
1, & m\geq n, \\ 
0, & n\geq m+1,
\end{array}
\right.  \label{29}
\end{equation}
and the ordered product of matrices, 
\begin{equation}
\overleftarrow{\prod_{l=n+1}^{m-1}}u_{l}^{}=\displaystyle\left\{ 
\begin{array}{cc}
u_{m-1}^{}u_{m-2}^{}\ldots u_{n+1}^{}, & m\geq n+2, \\ 
1, & m=n+1,
\end{array}
\right. .  \label{30}
\end{equation}
Because of Eqs.~(\ref{25}),~(\ref{6}), and~(\ref{23}) 
\begin{equation}
u_{n}^{}=u=w_{}^{\sigma },\qquad n\leq k-1,\qquad n\geq K+1,  \label{31}
\end{equation}
i.e., $u$ is a diagonal matrix independent on the regularization parameter 
$\lambda $ . Let us consider first $m\leq k$. Then by~(\ref{31}) we can
rewrite~(\ref{28}) in the form 
\[
u_{}^{-m}\widetilde{M}_{m,n}^{}-\theta (m\geq
n+1)u_{}^{-n-1}=u_{}^{-m'}\widetilde{M}_{m',n}^{}-\theta
(m'\geq n+1)u_{}^{-n-1}. 
\]
We see that both sides of this equality are independent either on $m$, or on 
$m'$; we denote them as $F_{n}$ and thus we get 
\begin{equation}
\widetilde{M}_{m,n}^{}=\theta (m\geq n+1)u_{}^{m-n-1}+u_{}^{m}F_{n}^{},
\qquad m\leq k.  \label{32}
\end{equation}
Now we chose in~(\ref{28}) $m'=k$ and substitute $\widetilde{M}_{k,n}$ in 
the r.h.s.\ using~(\ref{32}), then 
\begin{eqnarray*}
&&\widetilde{M}_{m,n}^{} =\theta (m\geq n+1)\theta (n\geq k)\overleftarrow
{\prod_{l=n+1}^{m-1}}u_{l}^{}+\\
&&\qquad +\theta (k\geq n+1)\left( \overleftarrow
{\prod_{l=k}^{m-1}}u_{l}^{}\right) u^{k-n-1}+ 
\left( \overleftarrow{\prod_{l=k}^{m-1}}u_{l}^{}\right) u_{}^{k}F_{n}^{},
\end{eqnarray*}
where $m\geq k$. Thus the second term also obeys condition $m\geq n+1$, so 
that taking~(\ref{25}) into account we can write 
\begin{equation}
\widetilde{M}_{m,n}^{}=\theta (m\geq n+1)\overleftarrow{\prod_{l=n+1}^{m-1}}
u_{l}^{}+\left( \overleftarrow{\prod_{l=k}^{m-1}}u_{l}^{}\right)
u_{}^{k}F_{n}^{},\qquad m\geq k.  \label{a33}
\end{equation}

Eqs.~(\ref{32}) and~(\ref{a33}) give the general solution of~(\ref{15}) 
for any $F_{n}$. In order to fix it, we use the two conditions formulated
above. First of all it is necessary to guarantee convergency for any $n$ of
the series 
$\sum_{m}\zeta^{-m}M_{m,n}=\sum_{m}(h\zeta)^{-m}\widetilde{M}_{m,n}$, 
where Eq.~(\ref{26}) was used. Let us consider first the sum from $-\infty$ 
to $k$. Using~(\ref{32}) we see that the sum of the first terms is finite 
due to $\theta$-function. The sum of the second terms in~(\ref{32}) is equal 
(up to a constant factor) to 
$\sum_{m=-\infty}^{k}u^{m}(h\zeta)^{-m}F_{n}$. Thanks to~(\ref{31}) this 
sum converges iff the first (second) row of matrix $F_{n}$ is equal to zero 
when $|w|<h$ ($1/|w|<h$, correspondingly). Thus the condition of convergency 
of this series can be written as 
\begin{equation}
\theta (h-|w|_{}^{\sigma })F_{n}^{}=0,\qquad n\in \hbox{\bbb Z},  \label{34}
\end{equation}
where the matrix $\theta$-function is defined in~(\ref{a22}). Let us consider
now the condition of convergency of the series $\sum_{m}\zeta^{-m}M_{m,n}$ 
at plus infinity. For this purpose we write Eq.~(\ref{a33}) for $m\geq K+1$ 
(see~(\ref{25}) and~(\ref{31})) as 
\begin{eqnarray}
&&\widetilde{M}_{m,n}^{} =u_{}^{m-1} \Biggl\{\theta (n\geq K)u_{}^{-n}+\theta
(K\geq n+1)u_{}^{-K}\overleftarrow {\prod_{l=n+1}^{K}}u_{l}^{}+  \nonumber \\
&&\qquad +u_{}^{-K}\left( \overleftarrow{\prod_{l=k}^{K}}u_{l}^{}\right)
u_{}^{k}F_{n}^{} \Biggr\} -\theta (n\geq m)u_{}^{m-n-1}.  \label{35}
\end{eqnarray}

By~(\ref{26}) the series 
$\sum_{m=K+1}^{\infty }(h\zeta)^{-m}\widetilde{M}_{m,n}$ must be convergent. 
The sum of the last terms is finite due to the $\theta$-function and the sum
of the first terms is convergent iff 
\begin{eqnarray}
&&\theta (|u|-h)\Biggl\{\theta (n\geq K)u_{}^{-n}+\theta (K\geq n+1)u_{}^{-K}
\overleftarrow{\prod_{l=n+1}^{K}}u_{l}^{}+  \nonumber \\
&&\qquad +u_{}^{-K}\left( \overleftarrow{\prod_{l=k}^{K}}u_{l}^{}\right)
u_{}^{k}F_{n}^{}\Biggr\}=0.  \label{37}
\end{eqnarray}
The conditions~(\ref{34}) and~(\ref{37}) determine $F_{n}$ uniquely. In 
order to get its explicit form, we have to consider the four regions of 
continuity of the matrices (\ref{a22}): 
\begin{eqnarray}
&&h >|w|,\qquad h>1/|w|,  \label{38} \\
&&|w| >h>1/|w|,  \label{39} \\
&&1/|w| >h>|w|,  \label{40} \\
&&|w| >h,\qquad 1/|w|>h.  \label{41}
\end{eqnarray}
Then $\widetilde{M}_{m,n}^{}$ is constructed explicitly using 
Eqs.~(\ref{32}) and~(\ref{a33}).

Let us introduce the (infinite) matrix column 
\begin{equation}
x _{m}^{}(w,\lambda )=\theta (m\geq k+1)\left( \overleftarrow
{\prod_{l=k}^{m-1}}u_{l}^{}(w,\lambda )\right) w_{}^{k\sigma }+\theta (k\geq
m)w_{}^{m\sigma },  \label{42}
\end{equation}
and row 
\begin{equation}
y _{n}^{}(w,\lambda )=\theta (n\geq K)w_{}^{-n\sigma }+\theta (K\geq
n+1)w^{-K\sigma }\overleftarrow{\prod_{l=n+1}^{K}}u_{l}^{}(w,\lambda ).
\label{a44}
\end{equation}
In what follows we also use 
\begin{equation}  \label{c44}
X_m^{}(w,\lambda,h)=h^{-m}_{}x_m^{}(w,\lambda),\qquad
Y_n^{}(w,\lambda,h)=h^{n}_{}y_n^{}(w,\lambda).
\end{equation}
Let $a$ denote a constant (i.e., independent on $m$ and $n$) matrix 
\begin{equation}
a(w,\lambda )=w^{-K\sigma }\overleftarrow{\prod_{l=k}^{K}}
u_{l}^{}(w,\lambda )w_{}^{k\sigma }.  \label{b44}
\end{equation}
Then, combining~(\ref{32}) and~(\ref{a33}), we get 
\begin{equation}
\widetilde{M}_{m,n}^{}=\theta (m\geq n+1)\overleftarrow{\prod_{l=n+1}^{m-1}}
u_{l}^{}+x _{m}^{}F_{n}^{}  \label{421}
\end{equation}
and instead of~(\ref{37}) we can write 
\begin{equation}
\theta (|w_{}^{\sigma }|-h)aF_{n}^{}=-\theta (|w_{}^{\sigma }|-h)y _{n}^{}.
\label{422}
\end{equation}

Let us consider first the region~(\ref{38}). Then $\theta (h-|w^{\sigma }|)=1$
and by~(\ref{34}) $F_{n}=0$, so that~(\ref{422}) is satisfied identically.
This means that in this region 
\begin{equation}
\widetilde{M}_{m,n}^{}=\theta (m\geq n+1)\overleftarrow{\prod_{l=n+1}^{m-1}}
u_{l}^{}.  \label{43}
\end{equation}
In the region~(\ref{39}) 
\[
\theta (h-|w_{}^{\sigma }|)=\frac{1-\sigma }{2},\qquad \frac{1-\sigma }{2}
F_{n}^{}=0, 
\]
where Eq.~(\ref{34}) was used and by~(\ref{422}) 
\begin{equation}
F_{n}^{}=-\frac{1+\sigma }{2a_{1,1}^{}}y _{n}^{}.  \label{a44'}
\end{equation}
Analogously in the region~(\ref{40}) we have that 
\begin{equation}
F_{n}^{}=-\frac{1-\sigma }{2a_{2,2}^{}}y _{n}^{}.  \label{45}
\end{equation}
Finally, in the region~(\ref{41}) $\theta (h-|w^{\sigma }|)=0$, thus 
Eq.~(\ref{34}) is satisfied identically and~(\ref{37}) takes the form 
\begin{equation}
F_{n}^{}=-a_{}^{-1}y _{n}^{}.  \label{46}
\end{equation}
All this enables us to write that 
\begin{equation}
\widetilde{M}_{m,n}^{}=\theta (m\geq n+1)\overleftarrow{\prod_{l=n+1}^{m-1}}
u_{l}^{}-x _{m}^{}\Gamma y _{n}^{},  \label{47}
\end{equation}
where we introduced the matrix $\Gamma$ independent on $m$ and $n$,
\begin{equation}
\Gamma =\left\{\begin{array}{rl}
0, & \quad h>|w|\quad{\rm and}\quad h>1/|w|, \\ &  \\ 
\displaystyle\frac{1+\sigma }{2a_{1,1}^{}}, & \quad |w|>h\quad{\rm and}\quad 
h>1/|w|, \\ &  \\ 
\displaystyle\frac{1-\sigma }{2a_{2,2}^{}}, & \quad h>|w|\quad{\rm and}\quad 
1/|w|>h, \\ &  \\ 
a_{}^{-1}, & \quad |w|>h\quad{\rm and}\quad 1/|w|>h.
\end{array}
\right.  \label{48}
\end{equation}

Let us mention that thanks to~(\ref{31}) we can rewrite Eqs.~(\ref{a44}) 
and~(\ref{46}) as 
\begin{equation}
F_{n}^{}=-\theta (n\geq k)w_{}^{-k\sigma }\left( \overleftarrow
{\prod_{l=k}^{n}}u_{l}^{}\right) _{}^{-1}-\theta (k\geq
n+1)w_{}^{-(n+1)\sigma },  \label{49}
\end{equation}
and then after some simple calculations we get instead of~(\ref{47})
\begin{equation}
\widetilde{M}_{m,n}^{}=-\theta (n\geq m)\left( \overleftarrow
{\prod_{l=m}^{n}}u_{l}^{}\right) _{}^{-1}  \label{50}
\end{equation}
in the region~(\ref{41}).

The results of the above construction can be summarized as the following

{\bf Theorem 1.} The extended resolvent $M(w,\lambda )$ of the 
$L$-operator~(\ref{5}) regularized by~(\ref{23}) exists, is unique and 
equals to 
\begin{eqnarray}
&&M_{m,n}^{}(w,\lambda ,h) =h_{}^{n-m} \theta (m\geq n+1)\overleftarrow
{\prod_{l=n+1}^{m-1}}u_{l}^{}(w,\lambda )-  \nonumber \\
&&\qquad -X _{m}^{}(w,\lambda )\Gamma (w,\lambda ,h)Y _{n}^{}(w,\lambda ),  
\label{51}
\end{eqnarray}
where $X _{m}$, $Y _{n}$, and $\Gamma $ are given 
in~(\ref{42}),~(\ref{a44}),~(\ref{c44}), and~(\ref{48}).

To prove the theorem we need to check first of all that the double series 
\begin{equation}
M(w,\lambda ,\zeta,\zeta',h)=\sum_{m,n=-\infty }^{+\infty }\zeta
_{}^{-m}\zeta_{}^{\prime n}M_{m,n}^{}(w,\lambda ,h),  \label{52}
\end{equation}
where~(\ref{26}) was taken into account, converge in the sense of 
Definition~1. Then it is necessary to prove that~(\ref{51}) obeys both
(regularized) equations~(\ref{14}), i.e., equations~(\ref{241}). 

The solution~(\ref{51}) by construction is the unique solution of the first 
equation in~(\ref{241}) for which the series 
$\sum_{m=-\infty}^{+\infty}\zeta^{-m}M_{m,n}(w,\lambda ,h)$ converge. 
Convergency of the series in $\zeta'$, as well as the second equation, are 
proved analogously. Both these equations easily follows from~(\ref{51}), if 
we notice that by~(\ref {42}) and~(\ref{a44}) 
\begin{eqnarray}
&&x _{m+1}^{}(w,\lambda ) =u_{m}^{}(w,\lambda )x_{m}^{}(w,\lambda ),
\label{53} \\
&&y _{m-1}^{}(w,\lambda ) =y _{m}^{}(w,\lambda)u_{m}^{}(w,\lambda ).
\label{54}
\end{eqnarray}
In other words $x _{m}$ and $y _{n}$ are solutions of the 
equation~(\ref{3'}) regularized by~(\ref{23}) and its dual. By means 
of~(\ref{c44}) we can also write these equations as 
\begin{equation}
L(w,\lambda)X(w,\lambda)=0,\qquad Y(w,\lambda)L(w,\lambda)=0.  \label{541}
\end{equation}
We see that formally $X$ and $Y$ are right and left annulators of operator 
$L$. The existence of these annulators does not contradict~(\ref{241}), i.e.,
the existence of the inversion of $L$ as both series 
$\sum_m\zeta^{-m}X_m(w,\lambda,h)$ and $\sum_n\zeta^{n}Y_n(w,\lambda,h)$ are 
divergent, so $X_m$ and $Y_n$ do not belong to the space mentioned in 
discussion of Eq.~(\ref{7}) and in Definition 1. The use of such quantities 
can be avoided if, say, in the region $|w|>h$, $1/|w|>h$ we use instead 
of~(\ref{51}) the equality 
\begin{equation}
M_{m,n}^{}=-h^{n-m}_{}\theta (n\geq m)\left( \overleftarrow
{\prod_{l=m}^{n}}u_{l}^{}(w,\lambda)\right) _{}^{-1},  \label{542}
\end{equation}
that follows from~(\ref{26}) and~(\ref{50}).

\section{Extended resolvent of the original operator}

In order to get the resolvent of the extended original operator~(\ref{5}) we
need to consider the behavior of~(\ref{51}) in the limit 
$\lambda\rightarrow0$. The existence of this limit depends on the 
regions~(\ref{38})--(\ref{41}). Indeed, the only origin of a singularity 
in~(\ref{51}) is matrix $\Gamma $, as follows from~(\ref{42}) 
and~(\ref{a44}). Its limits in the first three regions of~(\ref{48}) exist 
by~(\ref{b44}). Let
\begin{equation}
a(w)=\lim_{\lambda \rightarrow 0}a(w,\lambda )=w^{-K\sigma } \left
(\overleftarrow{\prod_{l=k}^{K}}u_{l}^{}(w)\right)w_{}^{k\sigma }.  
\label{a55}
\end{equation}
This expression is finite and nonzero for generic $w$. Zeroes of 
$a_{1,1}(w)$ and $a_{2,2}(w) $ if they exist in the corresponding regions 
give bound states of operator $L(w)$ and will be studied in the following 
publication. In the region~(\ref{41}) $a^{-1}(w,\lambda)$ has pole at 
$\lambda=0$, as follows from the last line of~(\ref{48}).

To describe the multiplicity of this pole we introduce $q(m,n)$, $m\leq n$, 
number of the degenerated matrices $u_{l}^{}(w)$ on the interval $[m,n]$, 
i.e., 
\begin{equation}
q(m,n)=\sum_{l=m}^{n}(1-\det u_{l}^{}(w)),  \label{581}
\end{equation}
which is independent on $w$ as $\det u_{l}^{}(w)$ equals either $0$, or $1$.
Let also  
\begin{equation}
Q=q(k,K).  \label{58}
\end{equation}
Then by~(\ref{23}) and~(\ref{b44}) we have that 
\begin{equation}
a_{}^{-1}(w,\lambda )=\frac{w^{-k\sigma }}{\lambda _{}^{Q}}\left( 
\overrightarrow{\prod_{l=k}^{K}}\left[ \lambda w(1-\det u_{l}^{})\frac
{1-\sigma }{2}+\tilde{u}_{l}^{}(w)\right] \right) w_{}^{K\sigma },
\label{56}
\end{equation}
where we introduce the matrices 
\begin{equation}
\tilde{u}_{l}^{}(w)=\left( 
\begin{array}{rr}
1/w & -r_{n}^{} \\ 
-s_{n}^{} & w
\end{array}
\right) \equiv w_{}^{-\sigma }-\left( 
\begin{array}{rr}
0 & r_{n}^{} \\ 
s_{n}^{} & 0
\end{array}
\right) ,\qquad n\in \hbox{\bbb Z},  \label{57}
\end{equation}
which are the inverses of $u_{n}(w)$ in the case where $\det u_{n}=1$ 
(cf.~(\ref{6})). From~(\ref{56}) it follows that $a^{-1}$ has pole of order 
$Q$ at $\lambda =0$ and we can write 
\begin{equation}
a_{}^{-1}(w,\lambda )=\sum_{j=0}^{Q}\frac{t_{}^{(j)}(w)}{\lambda _{}^{j}}
+O(\lambda ),\qquad \lambda \rightarrow 0.  \label{59}
\end{equation}
The residues are equal to 
\begin{equation}
t_{}^{(j)}(w)=w^{Q-k\sigma }\left( \overrightarrow{\prod_{l=k}^{K}} \tilde{u}
_{l}^{}(w)\right) _{}^{(Q-j)}w_{}^{K\sigma },  \label{60}
\end{equation}
where by definition 
\begin{eqnarray}
&&\left( \overrightarrow{\prod_{l=m}^{n}}\tilde{u}_{l}^{}(w)\right)
_{}^{(j)}=\sum_{m\leq l_{1}<\ldots <l_{j}\leq n}
\left(\prod_{i=1}^{j}(1-\det u_{l_{i}}^{}(w))\right)\times  \nonumber \\
&&\qquad\times\left[\overrightarrow{\prod_{l=m}^{n}}\tilde{u}_{l}^{}(w)
\right]^{}_{\tilde{u}_{l_{i}}= \frac{1-\sigma }{2},\atop i=1,\ldots ,j}
\label{61}
\end{eqnarray}
for any $m\leq n$.

By~(\ref{42}),~(\ref{c44}), and~(\ref{a44}) $X_{m}(w,\lambda )$ and 
$Y_{n}(w,\lambda )$ are polynomials in $\lambda $, so that we have the Laurent
expansion 
\begin{eqnarray}
&&M(w,\lambda ,h)=\widehat{M}(w,\lambda ,h)+
\sum_{j=1}^{Q}\frac{M_{}^{(j)}(w)} {\lambda _{}^{j}},\label{62}\\
&&|w|>h\quad{\rm and}\quad  1/|w|>h,  \nonumber
\end{eqnarray}
where $\widehat{M}$ is the regular part of the series. The residues can be 
calculated explicitly by Eqs.~(\ref{51}) and~(\ref{59}), but in order to 
work with objects belonging to the space mentioned in Definition 1 it is 
reasonable to use the representation~(\ref{542}). Then by means of 
notations~(\ref{581}) and~(\ref{61}) we get 
\begin{eqnarray}
&&M_{m,n}^{(j)}(w,h) =-h_{}^{n-m}\theta (n\geq m)\theta (q(m,n)\geq j)\times 
 \nonumber \\
&&\qquad \times w^{q(m,n)-j}\left( \overrightarrow{\prod_{l=m}^{n}}\tilde{u}
_{l}^{}(w)\right) _{}^{(q(m,n)-j)}.  \label{63}
\end{eqnarray}

Now the resolvent $M(w,h)$ of the original (extended) $L$-operator~(\ref{5})
can be defined as 
\begin{equation}
M(w,h)=\left\{ \begin{array}{cc}
M(w,\lambda ,h)\mid _{\lambda =0}^{}, & \quad h>|w|\quad {\rm or}\quad
h>1/|w|, \\ 
\widehat{M}(w,\lambda ,h)\mid _{\lambda =0}^{}, & \quad |w|>h\quad{\rm and}
\quad 1/|w|>h.
\end{array}
\right.  \label{64}
\end{equation}

Let us consider region $|w|>h$, $1/|w|>h$ in detail. Inserting the 
expression~(\ref{62}) into Eqs.~(\ref{242}) and~(\ref{243}) and using 
(\ref{64}) we derive that 
\begin{eqnarray}
&&L(w)M(w) =I+DM_{}^{(1)}(w),\quad M(w)L(w)=I+M_{}^{(1)}(w)D,\qquad   
\label{70} \\
&&L(w)M_{}^{(j)}(w) =I+DM_{}^{(j+1)}(w),\label{711}\\
&&M_{}^{(j)}(w)L(w)=I+M_{}^{(j+1)}(w)D,  
\qquad j =1,\ldots ,Q-1, \label{71} \\
&&L(w)M_{}^{(Q)}(w) =0,\quad M_{}^{(Q)}(w)L(w)=0.  \label{72}
\end{eqnarray}
Thus we see that in this region equations~(\ref{70}) defining the resolvent 
are modified in comparison with the standard Eqs.~(\ref{14}). In~[26] it was 
shown that a solution of~(\ref{14}) does not exist in this region. In order 
to study the properties of the residues $M^{(j)}$ we can use Hilbert 
identity 
\[
M(w,\lambda ')[L(w,\lambda ')-L(w,\lambda )]M(w,\lambda
)=M(w,\lambda )-M(w,\lambda ')
\]
that follows from~(\ref{241}). Taking into account~(\ref{242}) 
and~(\ref{243}) we can rewrite it in the form 
\begin{equation}
M(w,\lambda )-M(w,\lambda ')=(\lambda '-\lambda)
M(w,\lambda ')DM(w,\lambda ).  \label{73}
\end{equation}
Substituting $M(w,\lambda ')$ as in Eq.~(\ref{62}), we get in the limit 
$\lambda '\rightarrow 0$ that 
\begin{eqnarray}
&&M(w,\lambda )-M(w) =[\lambda M(w)-M_{}^{(1)}(w)]DM(w,\lambda ),\label{74}\\
&&M_{}^{(j)}(w) =[M_{}^{(j+1)}(w)-\lambda M_{}^{(j)}(w)]DM(w,\lambda ),
\label{75}
\end{eqnarray}
where $j=1,\ldots ,Q$, and we put by definition 
\begin{equation}
M_{}^{(Q+1)}(w)\equiv 0.  \label{76}
\end{equation}
Now we insert the expansion~(\ref{64}) in the Eqs.~(\ref{74}) 
and~(\ref{75}), and by~(\ref{74}) in the limit $\lambda \rightarrow 0$ we 
derive that
\begin{eqnarray}
&&M_{}^{(1)}(w)DM(w) =M(w)DM_{}^{(1)}(w),  \label{a77} \\
&&M_{}^{(j)}(w) =M(w)DM_{}^{(j+1)}(w)-M^{(1)}(w)DM_{}^{(j)}(w),
\label{78} \\
&&j =1,\ldots ,Q.  \nonumber
\end{eqnarray}
Then from~(\ref{75}) we have 
\begin{equation}
M_{}^{(j)}(w)=M_{}^{(j+1)}(w)DM(w)-M^{(j)}(w)DM_{}^{(1)}(w)  \label{79}
\end{equation}
(symmetric to~(\ref{78})) and 
\begin{eqnarray}
&&M_{}^{(j+1)}(w)DM_{}^{(l)}(w) =M_{}^{(j)}(w)DM_{}^{(l+1)}(w), \label{80} \\
&&M_{}^{(j)}(w)D\widehat{M}(w,\lambda ) =\frac{1}{\lambda }
M_{}^{(j+1)}(w)D[\widehat{M}(w,\lambda )-M(w)],  \label{81} \\
&&j,l =1,\ldots ,Q,  \nonumber
\end{eqnarray}
where~(\ref{64}) and~(\ref{76}) were used. By~(\ref{79}) we have 
$M^{(Q)}(w)=-M^{(Q)}(w)DM^{(1)}(w)$ and by~(\ref{81}) 
$M^{(Q)}(w)D\widehat{M}(w,\lambda )=0$, so that also $M^{(Q)}(w)DM(w)=0$ 
by~(\ref{64}). Then for $j=Q-1$ in~(\ref{79}) we have that 
$M^{(Q-1)}(w)=-M^{(Q-1)}(w)DM^{(1)}(w)$ and from~(\ref{81}) that 
\newline$M^{(Q-1)}(w)D\widehat{M}(w,\lambda )=0$, so that again 
$M^{(Q-1)}(w)DM(w)=0$. Continuing in this way we prove that 
\begin{eqnarray}
&&M_{}^{(j)}(w) =-M_{}^{(j)}(w)DM_{}^{(1)}(w),  \label{82} \\
&&M_{}^{(j)}(w) =-M_{}^{(1)}(w)DM_{}^{(j)}(w),  \label{83}
\end{eqnarray}
and 
\begin{eqnarray}
&&M_{}^{(j)}(w)DM(w) =0,\quad M(w)DM_{}^{(j)}(w)=0,  \label{84} \\
&&j =1,\ldots ,Q,  \nonumber
\end{eqnarray}
where~(\ref{83}) and the second set of equations in~(\ref{84}) is derived in 
an analogous way from~(\ref{73}), if $\lambda $ and $\lambda '$ are
interchanged. 

Now by~(\ref{80}) we get
\begin{equation}
M_{}^{(j)}(w)DM_{}^{(l)}(w)=M_{}^{(j-k)}(w)DM_{}^{(l+k)}(w),
\label{85}
\end{equation}
where $j,l=1,\ldots ,Q$, $k\geq 0$, $j-k\geq 1$, $l+k\leq Q+1$. If $j$ and $l
$ are such that $j-k\geq 1$ and $l+k=Q+1$, i.e., $l+j\geq Q+2$, then the
r.h.s.\ of~(\ref{85}) is equal to zero thanks to~(\ref{76}). On the
other side if $l+j\leq Q+1$ we can chose in~(\ref{85}) $k=j-1$ and then 
by~(\ref{82}) or~(\ref{83}) we get finally
\begin{eqnarray}
&&M_{}^{(j)}(w)DM_{}^{(l)}(w) =-\theta (Q+1\geq l+j)M_{}^{(l+j-1)}(w),
\label{86} \\
&&j,l =1,\ldots ,Q,  \nonumber
\end{eqnarray}
that replaces relations~(\ref{80}),~(\ref{82}), and~(\ref{83}).

This concludes the construction of the resolvent $M_{m,n}(w,h)$ of the 
extended $L$-operator~(\ref{5}). As we have seen, this resolvent is 
discontinuous at $|w|=h$ and $|w|=1/h$. In a forthcoming paper we show 
that study of this discontinuity leads us to modification of the Jost 
solutions and spectral data, corresponding to the case of the discrete 
potential~(\ref{2}).

{\bf Acknowledgments.} This work is supported in part by RFBR under Grant
No.~96-01-00344.

\end{document}